# Elastic wave propagation in simple-sheared hyperelastic materials with different constitutive models

Linli Chen, Zheng Chang*, Taiyan Qin

College of Science, China Agricultural University, Beijing 100083, China

**Abstract**

We investigate the elastic wave propagation in various hyperelastic materials which subjected to simple-shear deformation. Two compressible types of three conventional hyperelastic models are considered. We found pure elastic wave modes can be obtained in compressible neo-Hookean materials constructed by adding a bulk strain energy term to the incompressible strain energy function. Whereas for the compressible hyperelastic models which are reformulate into deviatoric and hydrostatic parts, only quasi modes can propagate, with abnormal ray directions can be observed for longitudinal waves. Moreover, the influences of material constants, material compressibility and external deformations on the propagation and refraction for elastic waves in these hyperelastic models are systematically studied. Numerical simulations are carried out to validate the theoretical results. This investigation may open a promising route for the realization of next generation metamaterials and novel wave manipulation devices.

**Keyword**: elastic wave; hyperelastic; stain energy function; compressible.

* Author to whom correspondence should be addressed. Electronic mail: changzh@cau.edu.cn (Z. Chang).

# 1 Introduction

Soft materials such as elastomers, gels and many biological tissues usually exhibit rich and complex static and dynamic behaviors when subjected to finite deformation. To describe the nonlinear mechanical behavior of such materials, hyperelasticity is usually employed. Many hyperelastic models (or Strain Energy Functions, SEFs) (Simo and Pister, 1984; Arruda and Boyce, 1993; Gent, 1996; Hartmann and Neff, 2003) are proposed with the quality of the fit to experimental data (Ogden, 1972; Yeoh, 1993; Boyce and Arruda, 2012), and hereafter applied for theoretical and numerical analyses. In recent days, soft materials with hyperelastic SEFs have drawn considerable attention in elastodynamics. In particular, by virtue of the high sensitivity to deformations and the remarkable capability of reversible structural instability of soft materials, soft metamaterials or soft phononic crystals (Bertoldi and Boyce, 2008; Auriault and Boutin, 2012; Wang and Bertoldi, 2012; Shim et al., 2015) with tunable or adaptive properties have been demonstrated for wave application. Moreover, a hyperelastic transformation theory (Norris and Parnell, 2012; Parnell, 2012; Chang et al., 2015; Liu et al., 2016) has been reported, providing homogeneous soft materials with certain SEFs can be utilized to manipulate elastic wave paths. In all these works, it is generally revealed that subtle difference on hyperelastic models may cause significant, or even subversive distinctions on the performance of the soft devices. Therefore, comprehensively understanding of the mechanical behaviors for different SEFs is essential, especially for some applications where high-precision is required.

Hyperelastic materials are usually considered to be incompressible. However, compressible hyperelastic model is indispensable in considering the longitudinal wave motion in elastodynamic problems. There are a couple of approaches to extend an incompressible hyperelastic model to a compressible form (Boyce and Arruda, 2012) . It has been demonstrated (Ehlers and Eipper, 1998) these two versions of the SEFs have significantly different static behaviors, especially in case the volume of material changes a lot. However, how these SEFs relate to their dynamic properties is still elusive.

Moreover, a previous work (Chang et al., 2015) has proposed a feasible method to separate longitudinal and shear elastic waves with a simple-sheared neo-Hookean solid. Considering the diversity of the hyperelastic models, there remains an unmet need for understanding the ability of wave-mode separation for different hyperelastic models.

To address above issues, in this work, we focus on the propagation and refraction of elastic waves in simple-sheared hyperelastic materials. In the framework of Small-on-Large theory (Ogden, 2007), the dynamic behaviors of two different compressible types of the three conventional incompressible hyperelastic models are considered. We show that pure elastic wave modes may propagate in a compressible neo-Hookean model constructed by adding a bulk strain energy term to a conventional incompressible SEF. For a compressible model reformulate from a conventional incompressible SEF into a deviatoric part and a hydrostatic part, only quasi wave modes exist. In considering the refraction of elastic waves normally incidence on a plane interface from an undeformed hyperelastic material to a pre-deformed hyperelastic material, significant difference can be observed in the refraction angle of the longitudinal wave between the two versions of SEFs. Both theoretical analysis and numerical simulations are carried out to confirm each other. The paper is organized as follows. Some theoretical backgrounds are provided in Sections. 2 and 3, in which the approaches to extend the incompressible SEFs to the compressible ones are briefly reviewed in Section. 2. Whereas the Small-on-Large theory which describes linear wave motion propagates in a finitely deformed hyperelastic material is recalled in Section. 3. Moreover, the elastic wave propagation and refraction in a simple-sheared hyperelastic material with different SEFs are shown in Section. 4, with numerical validations provided in Section.5. At the end, discussion on our results and avenues for future work is given in Section. 6.

**2 Strain energy functions for compressible hyperelastic materials**

In this section, we briefly review the two approaches (Boyce and Arruda, 2012) to expand an incompressible SEF to a compressible one.

Approach I is to simply add a bulk strain energy term $W_B$ to an existing incompressible isotropic SEF $W_C$. Thus, the compressible SEF can be expressed as

$$W_1 = W_C(I_1, I_2, J) + W_B(J), \tag{1}$$

where $I_1$ and $I_2$ are the first and second invariant of the right Cauchy-Green tensor, respectively. $J = \det(\mathbf{F})$ is the volumetric ratio. $F_{ij} = \partial x_i / \partial X_j$ denotes the deformation gradient, in which $X_j$ and $x_i$ are the coordinates in the initial and the current configurations, respectively.

Approach II is to apply a multiplicative decomposition on the Cauchy-Green deformation tensor, and reformulate the incompressible SEF into a deviatoric one. A hydrostatic strain energy term $W_H$ is then added to extend the SEF into a compressible form, namely

$$W_2 = W_D(\bar{I}_1, \bar{I}_2) + W_H(J), \tag{2}$$

where $\bar{I}_1 = J^{-1} I_1$ and $\bar{I}_2 = J^{-2} I_2$ are the invariants of the deviatoric stretch tensor.

In Eqs. (1) and (2), bulk stain energy terms $W_B$ or $W_H$ can be selected from some empirical formulas, such as $W_{H1}(J) = B(\ln J)^2 / 2$, $W_{H2}(J) = B((J^2 - 1)/2 - \ln J)/2$, $W_{H3}(J) = B(J-1)^2 / 2$, and $W_{H4}(J) = B\{\cosh[\alpha(J-1)] - 1\}/\alpha^2$, which are provided in former literatures (Bischoff et al., 2001; Hartmann and Neff, 2003). In these formulas, $B$ and $\alpha$ are material parameters, which can be determined from conditions (Simo and Pister, 1984) of $W(\mathbf{F} = \mathbf{I}) = 0$, $\partial W(\mathbf{F} = \mathbf{I})/\partial \mathbf{F} = 0$ and $\partial^2 W(\mathbf{F} = \mathbf{I})/\partial \mathbf{F} \partial \mathbf{F} = \lambda \delta_{ij}\delta_{kl} + \mu \delta_{ik}\delta_{jl} + \mu \delta_{il}\delta_{jk}$.

In the following, specific examples are identified for future reference. All the SEFs proposed can be referred from previous publications. For simplicity, we consider all the SEFs are in their two-dimensional (2-D) forms.

A compressible model of neo-Hookean SEF constructed by Approach I can be written as (Ogden, 1997)

$$W_{NH1} = \frac{\mu}{2}(I_1 - 2) - \mu \ln(J) + \frac{\lambda}{2}(J-1)^2, \tag{3}$$

where $\lambda$ and $\mu$ are the first and the second *Lamé* constants. The corresponding form constructed by Approach II can be found as (Ehlers and Eipper, 1998)

$$W_{NH2} = \frac{\mu}{2}(\bar{I}_1 - 2) + \frac{\kappa}{2}(\ln J)^2, \tag{4}$$

where $\kappa = \lambda + \mu$ is the bulk modulus.

Similarly, two typical compressible forms of Arruda-Boyce model can also be found in previous contributions (Boyce and Arruda, 2012) and (Kaliske and Rothert, 1997), which can be written as

$$W_{AB1} = C_1[\frac{1}{2}(I_1 - 2) + \frac{1}{20N}(I_1^2 - 4) + \frac{11}{1050N^2}(I_1^3 - 8) + \frac{19}{7000N^3}(I_1^4 - 16)$$
$$+ \frac{519}{673750N^4}(I_1^5 - 32)] - \mu \ln(J) + \frac{\lambda}{2}(J-1)^2, \tag{5}$$

and

$$W_{AB2} = C_1[\frac{1}{2}(\overline{I}_1 - 2) + \frac{1}{20N}(\overline{I}_1^2 - 4) + \frac{11}{1050N^2}(\overline{I}_1^3 - 8) + \frac{19}{7000N^3}(\overline{I}_1^4 - 16)$$
$$+ \frac{519}{673750N^4}(\overline{I}_1^5 - 32)] + \frac{\lambda + \mu}{2}\left(\frac{J^2-1}{2} - \ln J\right), \tag{6}$$

where $C_1 = \mu/(1 + 2/5N + 44/175N^2 + 152/875N^3 + 834/67375N^4)$ and $N$ is a material constant denotes a measure of the limiting network stretch. Note form of $C_1$ here is only for 2-D SEFs. When $N \to \infty$, Arruda-Boyce models degenerate to the neo-Hookean ones.

Another commonly utilized hyperelastic model is Gent model (Gent, 1996). Here we only consider its compressible form constructed by Approach I, which can be referred from an earlier work (Bertoldi and Boyce, 2008), namely

$$W_{G1} = -\frac{\mu J_m}{2}\ln(1 - \frac{I_1 - 2}{J_m}) - \mu \ln J + (\frac{\lambda}{2} - \frac{\mu}{J_m})(J-1)^2, \tag{7}$$

where $J_m$ is a material constant related to the strain saturation of the material. Similar as Arruda-Boyce model, Gent model degenerates to neo-Hookean model when $J_m \to \infty$.

### 3 Small-on-Large wave motion in hyperelastic materials

The Small-on-Large theory provides an ideal platform to investigate linear waves propagating in finite-deformed hyperelastic materials. For a hyperelastic solid with certain SEF, the equilibrium equation of the finite deformation can be written as

$$(A_{ijkl}U_{l,k})_{,i} = 0, \tag{8}$$

where $U_i$ denotes the displacement, $A_{ijkl} = \partial^2 W/\partial F_{ji}\partial F_{lk}$ are the components of the fourth-order elastic tensor expressed in the initial configuration. Further, the governing equation of the linear wave motion $u_i$ superimposed on the finite deformation $U_i$ can be written as (Ogden, 2007)

$$(A_{0i'jk'l}u_{l,k})_{,i} = \rho_0 \ddot{u}_j, \tag{9}$$

where $A_{0i'jk'l} = J^{-1}F_{i'i}F_{k'k}A_{ijkl}$ and $\rho_0 = J^{-1}\rho$ are the elastic tensor and mass density of the current configurations, respectively. $\rho$ is the density of the initial material.

For a homogenously deformed hyperelastic material, incremental plane waves can be expressed in the form of (Ogden, 2007)

$$u_i = m_i f\left(kl_j \cdot x_j - \omega t\right), \tag{10}$$

where **m** is a unit polarization vector, $f$ denotes a twice continuously differentiable function, **l** is the unit vector in the wave direction, $\omega$ is the angular frequency of the elastic waves and $k$ is the wave number. Inserting Eq. (10) into Eq. (9), the Christoffel equation can be obtained as

$$A_{0i'jk'l}l_{i'}l_{k'}m_l = c^2 \rho_0 m_j, \tag{11}$$

where $c = \omega/k$ is the velocity of the wave. By solving the eigenvalue problem of Eq. (11), phase velocities ($V_P$ and $V_S$) and polarization angles ($\psi_P$ and $\psi_S$) of the P- and S-waves propagate in pre-deformed hyperelastic material can be obtained. The phase-slowness of the P- and S-waves can thus be yield as $S_{P,S} = 1/V_{P,S}$. Moreover, at fixed frequency, slowness curves that give $S_{P,S}$ as a function of the wave direction **l** can be plotted. Further, the ray velocities of the P- and S-waves can be obtained by the rule of $\mathbf{V}^r_{P,S} \cdot \mathbf{l} = \mathbf{V}_{P,S}$ (Auld, 1973). The slopes of the ray directions can be expressed as functions of the wave direction as

$$\tan \varphi_{P,S} = \frac{s_{P,S} \sin \varphi - (ds_{P,S}/d\theta) \cos \varphi}{s_{P,S} \cos \varphi + (ds_{P,S}/d\theta) \sin \varphi}, \tag{12}$$

where $\varphi = \arctan(l_y/l_x)$ is the wave direction, $\varphi_P = \arctan(V^r_{Py}/V^r_{Px})$ and $\varphi_S = \arctan(V^r_{Sy}/V^r_{Sx})$ denote the directions of the rays.

To consider the refraction of elastic waves on a plane interface between two homogeneously pre-deformed hyperelastic materials, the refraction angles of the elastic waves can be obtained by the graphical method proposed by a previous literature (Rokhlin et al., 1985). In this work, we particularly consider the refraction of elastic waves normally incidence on a plane interface from an un-deformed hyperelastic material to a pre-deformed hyperelastic material. In this case, the refraction angles are $\theta_{P,S} = \varphi_{P,S}$ for any given $\varphi$. We define the separation angle $\Delta\theta = |\theta_S - \theta_P|$ to examine the difference between the P- and S-waves.

**4 Elastic wave propagation in simple-sheared hyperelastic materials**

We first consider the propagation of P- and S-waves in a hyperelastic material in a homogeneously simple-shear deformation state. For the plane-strain problem, homogeneous displacement can be defined by the deformation gradient $F_{11} = F_{22} = 1$, $F_{12} = 0$ and $F_{21} = \tan \gamma$, where $\gamma$ denotes the shear angle of the simple-sheared deformation. Note that in this case the volume of the material is unchanged, i.e. $J = 1$. Moreover, we further consider the refraction of elastic waves normally incidence on a simple-sheared hyperelastic material, as a schematic of the problem illustrated in Fig. 1. In this case, plane interface is parallel to the shear direction. Therefore, the refraction angles of the P- and S-waves, which also denote the directions of ray velocities, can be yield from the slowness curves at $\varphi = 0$.

To investigate the effect of two compressible versions of neo-Hookean models ($W_{NH1}$ and $W_{NH2}$) on the elastic wave propagation, the slowness curves have been plotted in Figs. 2 (a) and (b). For both SEFs, the normalized initial parameters are set to be $\lambda=4$, $\mu=1$ and $\rho=1$, whereas the shear angle is set to be $\gamma = \arctan(1/3)$. It can be seen the slowness curves of the S-waves turn out to have a similar shape. For P-waves, however, the slowness curves exhibit different orientations, leading to a significant difference in the directions of the ray velocities. As is shown in Fig. 2 (c), the ray direction of P-wave in material $W_{NH1}$ is positive valued when the wave direction is in a range of $\varphi \in [0, \pi/4]$. In contrast, ray direction is negative for the case of $W_{NH2}$. Moreover, the polarization angles for the wave directions in a range of $\varphi \in [0, \pi/4]$ have been plotted in Fig.2 (d), showing that elastic waves in material $W_{NH1}$ are always in pure modes, which means the polarization direction are parallel (or normal) to the propagation directions of the P- (or S-) wave. On the contrary, quasi elastic waves can be observed in material $W_{NH2}$.

To examine the influences of material parameters on elastic wave behavior in these hyperelastic models, slowness curves of the three hyperelastic materials $W_{NH1}$, $W_{G1}$ and $W_{AB1}$ with different material constants are demonstrated in Figs. 3 (a) and (b). The normalized initial parameters and the shear angle are also set to be $\lambda=4$, $\mu=1$, $\rho=1$ and $\gamma = \arctan(1/3)$. For $W_{NH1}$, $W_{G1}(J_m = 10)$ and $W_{AB1}(N=1)$, considerable difference can be perceived in the slowness curves, as demonstrated in Fig. 3 (a). However, for larger material constants, e.g. $J_m = 20$ and $N=3$, the slowness curves almost coincide with each other. For the refraction problem described in Fig. 1, the influence of compressibility on the three models has been plotted in Fig. 3 (c). Here, the normalized initial parameters and the shear angle are set to be $\mu=1$, $\rho=1$ and $\gamma = \arctan(1/3)$. $\lambda/\mu$ is introduced to represent the compressibility of the material. With the increase of the material compressibility, the refraction angles of the P-waves significantly decrease, while that of the S-waves keeps almost unchanged. Among the three models, $W_{NH1}$ exhibits the largest separation angle. However, for $W_{AB1}(N=1)$, the corresponding separation angle is much smaller, and the two wave paths almost coincide in case the material is nearly incompressible. The refraction angle of the P-wave progressively reduces for increasing values of $\lambda/\mu$. Moreover, the influence of material constants on the refraction angles of the P- and S-waves has been plotted in Fig.3 (d). The normalized initial parameters and the shear angle are set to be $\lambda=4$, $\mu=1$, $\rho=1$ and $\gamma = \arctan(1/3)$. For large material constants, the separation angles of the three SEFs tend to be consistent, because in this case, both $W_{G1}$ and $W_{AB1}$ degenerate into $W_{NH1}$. However for small material constants, considerable difference on the separation angles can be perceived for different SEFs.

Correspondingly, the influences of material constants and material compressibility on elastic wave behavior in materials $W_{NH2}$ and $W_{AB2}$, are demonstrated in Fig. 4. It is worthy of note in Figs. 4 (a) and (b), the orientation of the S-waves are in a same manner, in contrast with that shown in Figs. 3 (a) and (b). This reveals the ray directions of the P-waves are characterized by the approaches selected to construct the compressible SEFs. Considering the refraction problem, abnormal ray directions lead to negative refractions of P-waves for both SEFs, which is distinct from that of the SEFs constructed by Approach I. This will consequently result in a larger separation angle of $W_{AB2}$ than that of $W_{NH2}$, as shown in Fig. 4 (d). It also should be noted that the property of pure mode propagation is a unique feature for $W_{NH1}$, whereas not ubiquitous for the compressible SEFs constructed by Approach I.

To study the influence of shear angle $\gamma$ on the refraction properties of different

SEFs, the separation angles are plotted with respect to the shear angle of the simple-shear deformation, as shown in Fig. 5. The normalized initial parameters for all the SEFs are $\lambda=4$, $\mu=1$ and $\rho=1$. The material constants are set to be $J_m=10$ for $W_{G1}$, and $N=1$ for $W_{AB1}$ and $W_{AB2}$. In this figure, the linear property of neo-Hookean materials, $W_{NH1}$ and $W_{NH2}$, on the capacity of bearing shear deformation is clearly demonstrated. With the increases of the shear angle, the SEFs constructed by the different approaches exhibit diverse manner. Two possible wave separation modes can be distinguished, which are bounded by the lines respect to $W_{NH1}$ and $W_{NH2}$. Obviously, this "mode separation" is made by the distinguish refraction properties of P-waves.

## 5 Numerical Validations

To validate the theoretical results in Section. 4, numerical simulations have been performed by a two-step finite element model using the software COMSOL Multiphysics. In the first step, the finite deformation of the hyperelastic material is calculated with the solid mechanics module. After which, the deformed geometry configuration, together with the deformation gradient $\mathbf{F}$, is imported into the second step to simulate the linear elastic wave motion governed by Eq. (9). In this step, the weak form PDE module is applied to deal with the asymmetry of the elastic tensor.

Consider a square hyperelastic material with the side length $l=0.12\,\text{m}$ and the material parameters $\lambda=4.32\,\text{MPa}$, $\mu=1.08\,\text{MPa}$ and $\rho=1050\,\text{Kg}/\text{m}^3$. Apply the displacement field $U_y=(x+0.6)/30\,\text{m}$ to the model so that a homogeneous simple-sheared deformation can be obtained where the shear angle $\gamma$ satisfies $\tan\gamma=1/3$. As illustrated in Fig. 1, on the left side of the square, a P-wave beam and an S-wave beam are normally imported at the same position. Both the incident waves are set the maximum amplitude $u=1\times10^{-3}\,\text{m}$, with the angular frequencies of the S-wave and the P-wave are $\omega_S=0.3\,\text{MHz}$ and $\omega_P=12.9\,\text{MHz}$, respectively.

In general, the results of the numerical simulations are in excellent agreement with those obtained from theoretical analyses. The total displacement fields of the P- and S-waves propagate in material $W_{NH1}$ and $W_{NH2}$ are illustrated in Figs. 6 (a) and (b), respectively. Between these two figures, it is clearly seen that the S-waves have the similar refraction angles, whereas the P-waves exhibit opposite behavior. To better observe quasi wave modes in material $W_{NH2}$, the x- and y-components of the displacement field is illustrated in Figs. 6 (c) and (d), respectively. Small x- (or y-) displacement on the S- (or P-) wave path (shown in the insets in Figs. 6 (c) and (d) ) clearly demonstrates the polarization. As expected, no similar result can be observed in material $W_{NH1}$.

The distribution of the displacements on the right boundary of the materials (the green line in Figs. 6 (a) and (b)), are illustrated in Fig. 7. With an auxiliary line denoting the position of the incident wave, all the P- and S- wave fields shown in Figs. 7 (a) and (b) are distributed on the right side of the line. On the contrary, as is shown in Figs. 7 (c) and (d), the P- and S-wave fields are distributed in different sides, indicating that the negative refraction is a common feature for the SEFs constructed by Approach II. Moreover, it can be perceived in Fig. 7 (a), small perturbations of the wave fields exist for $W_{AB1}$ and $W_{G1}$ for small material constants. The perturbations reduce to the same level as that of $W_{NH1}$ for larger material constants, as shown in Fig. 7 (b). These findings recall the results in Section. 4 that the pure mode propagation property is unique for $W_{NH1}$, and once again confirmed the similar refraction behavior among $W_{NH1}$, $W_{AB1}$ and $W_{G1}$ with large material constants. For the SEFs constructed by Approach II, however, the perturbations always exist regardless of any SEF and material constants, as shown in Figs. 7 (c) and (d).

## 6 Discussion and Conclusion

In this work, we have investigated the propagation and refraction of elastic waves in pre-deformed hyperelastic materials with different SEFs. The above theoretical and numerical analyses show significant differences between the models constructed by the two SEF extension approaches. The variety of qualitatively different wave propagation behaviors gives us the opportunity to design novel architected materials with exotic functionalities.

In particular, we discover $W_{NH1}$ exhibits pure mode propagation of elastic waves when the material is subjected to pre-deformation. Such a behavior is anomalous for other hyperelastic materials and is beneficial for manipulation of elastic waves on the basis of hyperelastic transformation theory (Parnell, 2012; Parnell et al., 2012; Chang et al., 2015). It is also worthy of note $W_{AB1}$ and $W_{G1}$ with relatively large material constants have similar wave behaviors as that of $W_{NH1}$, which can also be applied to approximately manipulate S-waves.

On the separation of P- and S-waves, SEFs constructed by Approach I and II have their own merits. The eliminating of dealing with polarization makes $W_{NH1}$ prominent on the transmission and reception of elastic waves without energy loss. On the other hand, SEFs constructed by Approach II also have their advantage of large separation angles, which is crucial especially for some small scale applications. The effects of material compressibility, material constant and shear angle on the separation angle were also investigated. The results demonstrated that for the SEFs constructed by Approach I, more compressible material leads to smaller separation angle, which is on the contrary to the SEFs constructed by Approach II. Moreover, for all the SEFs, larger wave separation angles can be achieved with smaller material constants and larger shear angles.

This work also promotes an interesting and challenging direction on the material fabrication. With the development of modern chemical synthesis process (Chen et al., 2017), more elaborate theory and method are needed to precisely fabricate a material with a certain SEF as criteria. Only in this way can we freely utilize the distinguishing features among different SEFs.

**Acknowledgement**

This work was supported by the National Natural Science Foundation of China (Grant No. 11602294) and the Fundamental Research Funds for the Central Universities (Grant Nos. 2017QC056 and 2017LX002).

**Figure Captions**

**Fig. 1.** Schematic diagram of elastic waves propagating and refraction in a hyperelastic material under simple-shear. The black and orange frames denote the initial and deformed configurations, respectively. The material coordinates are denoted by orange grids. $\gamma$ is the shear angle. The blue and red lines represent the paths of P- and S-waves, respectively. $\theta_P$ and $\theta_S$ represent the refractive angles of P- and S-waves, respectively. In this case, the wave directions in the material is $\varphi = 0$.

**Fig. 2.** Elastic waves propagating in simple-sheared neo-Hookean materials with the SEFs of $W_{NH1}$ and $W_{NH2}$. (a) The slowness curves of the elastic waves propagating in the neo-Hookean material with SEF of $W_{NH1}$. $\varphi$ denotes the wave direction. The arrows denote the directions of the ray velocities. (b) The same as (a), but for the SEF of $W_{NH2}$. (c) The ray directions of the elastic waves with the wave direction of $\varphi \in [0, \pi/4]$. (d) The polarization angles of the elastic waves with the wave direction of $\varphi \in [0, \pi/4]$. For both SEFs, the normalized initial parameters are $\lambda=4$, $\mu=1$ and $\rho=1$, the shear angle is $\gamma = \arctan(1/3)$.

**Fig. 3.** Elastic waves propagating and refraction in simple-sheared hyperelastic materials with the SEFs of $W_{NH1}$, $W_{G1}$ and $W_{AB1}$. (a) The slowness curves of the elastic waves propagating in the hyperelastic materials with the SEFs of $W_{NH1}$, $W_{G1}$ ($J_m = 10$) and $W_{AB1}$ ($N = 1$). (b) The same as (a), but for $W_{NH1}$, $W_{G1}$ ($J_m = 20$) and $W_{AB1}$ ($N = 3$). In both (a) and (b), the normalized initial parameters are $\lambda=4$, $\mu=1$ and $\rho=1$, the shear angle is $\gamma = \arctan(1/3)$. (c) The refraction angles of the elastic waves with different material compressibility $\lambda/\mu \in (0.5, 10]$ for $W_{NH1}$, $W_{G1}$ ($J_m = 10$) and $W_{AB1}$ ($N = 1$), respectively. The initial mass density is $\rho=1$, the shear angle is $\gamma = \arctan(1/3)$ and the wave direction is $\varphi = 0$. (d) The angle between refracted P- and S-waves with different material constants $J_m$ and $N$. The normalized initial parameters are $\lambda=4$, $\mu=1$ and $\rho=1$, the shear angle is $\gamma = \arctan(1/3)$, and the wave direction is $\varphi = 0$.

**Fig. 4.** Elastic waves propagating and refraction in simple-sheared hyperelastic materials with the SEFs of $W_{NH2}$ and $W_{AB2}$. (a) The slowness curves of the elastic waves propagating in the hyperelastic materials with the SEFs of $W_{NH2}$ and $W_{AB2}$ ($N=1$). (b) The same as (a), but for $W_{NH1}$ and $W_{AB2}$ ($N=3$). In both (a) and (b), the normalized initial parameters are $\lambda=4$, $\mu=1$ and $\rho=1$, the shear angle is $\gamma = \arctan(1/3)$. (c) The refraction angles of the elastic waves with different material compressibility $\lambda/\mu \in (0.5, 10]$ for $W_{NH2}$ and $W_{AB2}$ ($N=1$). The initial mass density is $\rho=1$, the shear angle is $\gamma = \arctan(1/3)$ and the incident angle is $\varphi = 0$. (d) The angle between refracted P- and S-waves with different material constant $N$. The normalized initial parameters are $\lambda=4$, $\mu=1$ and $\rho=1$, the shear angle is $\gamma = \arctan(1/3)$, and the incident angle is $\varphi = 0$.

**Fig. 5.** The wave separation angle between refracted P- and S-waves with respect to different shear angle $\gamma$. The shaded areas I and II represent two possible wave separation modes. The normalized initial parameters are $\lambda=4$, $\mu=1$ and $\rho=1$, the shear angle is $\gamma = \arctan(1/3)$, and the wave direction is $\varphi = 0$. $J_m=10$ for $W_{G1}$, and $N=1$ for $W_{AB1}$ and $W_{AB2}$.

**Fig. 6.** Displacement fields in hyperelastic materials under simple shear. (a) Total displacement field in simple-sheared neo-Hookean material with the SEF of $W_{NH1}$. (b) The same as (a), but for $W_{NH2}$. Figs. (c) and (d) are the corresponding displacement components $u_x$ and $u_y$ with Fig. (b).

**Fig. 7.** Total displacement on auxiliary segments (green lines in Fig. 6) for different cases. The purple lines denote the position of the horizontal incident waves.

Fig.1

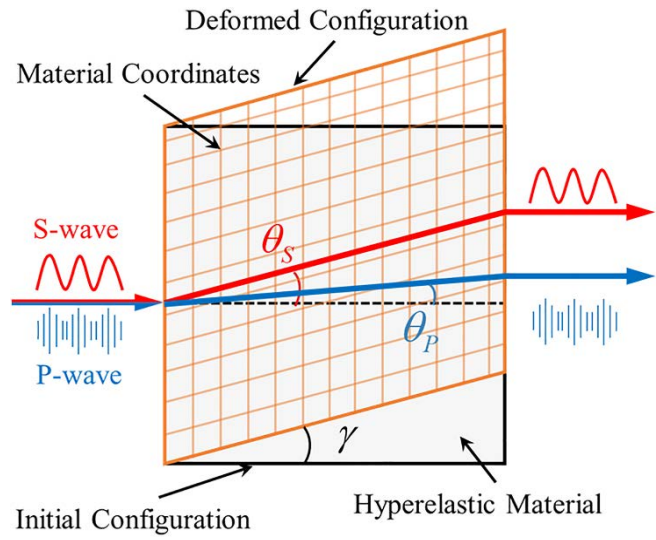

Fig.2

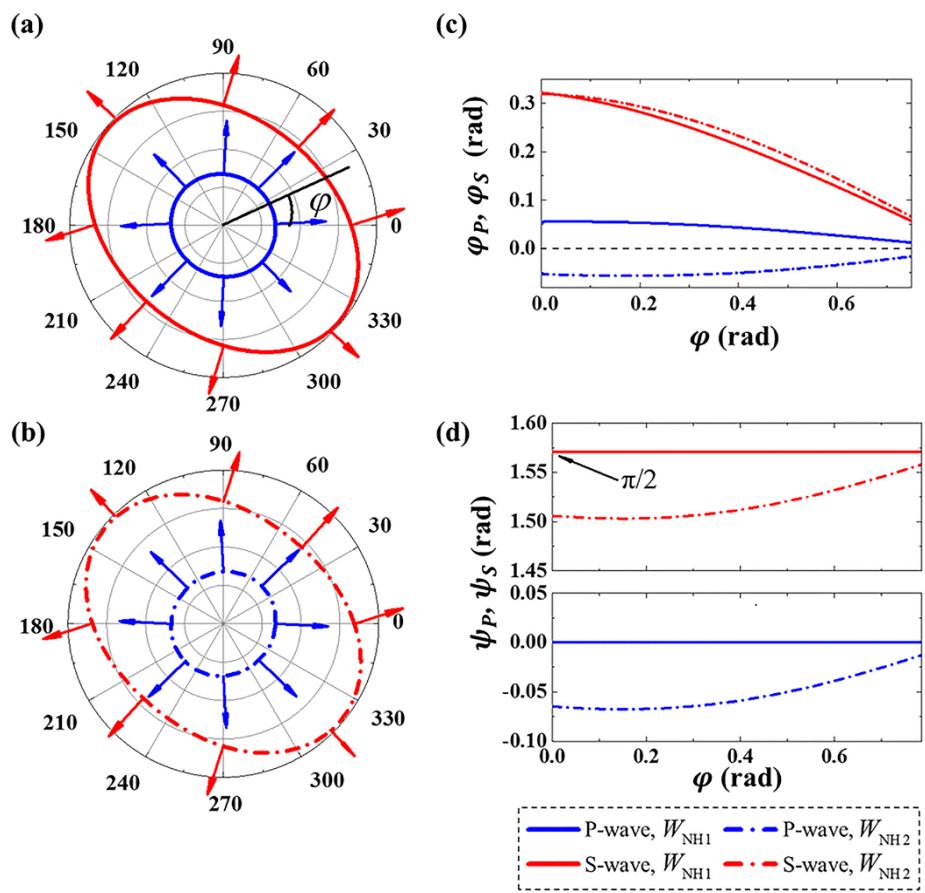

Fig.3

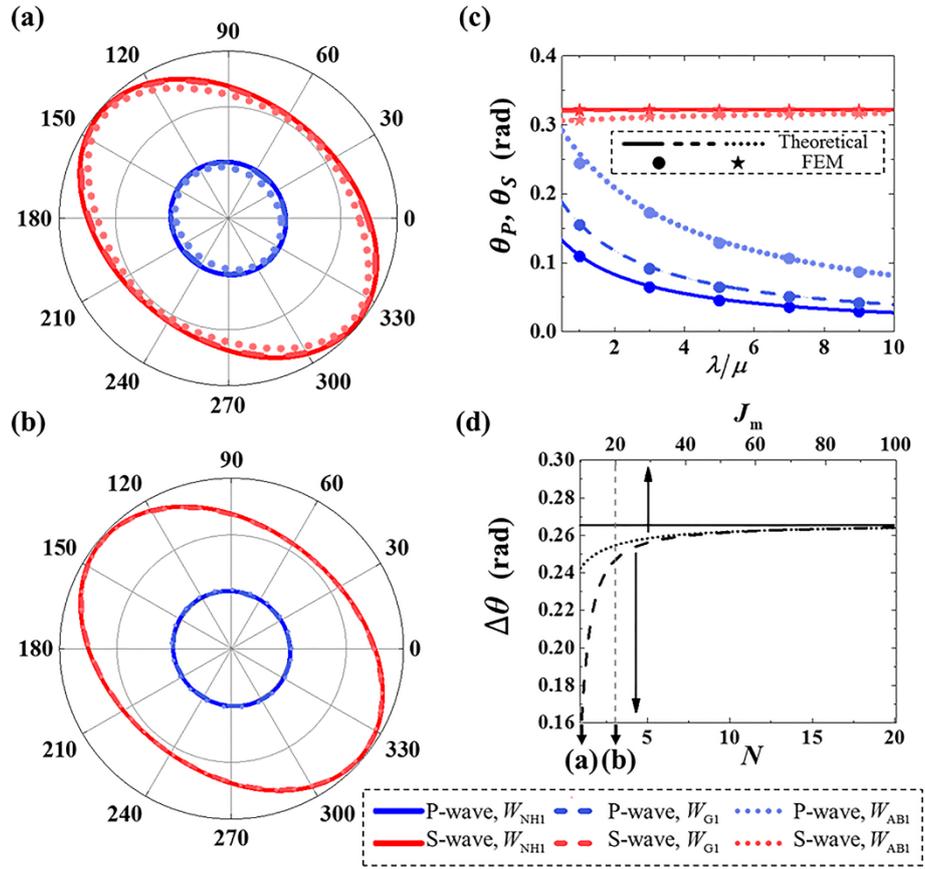

Fig.4

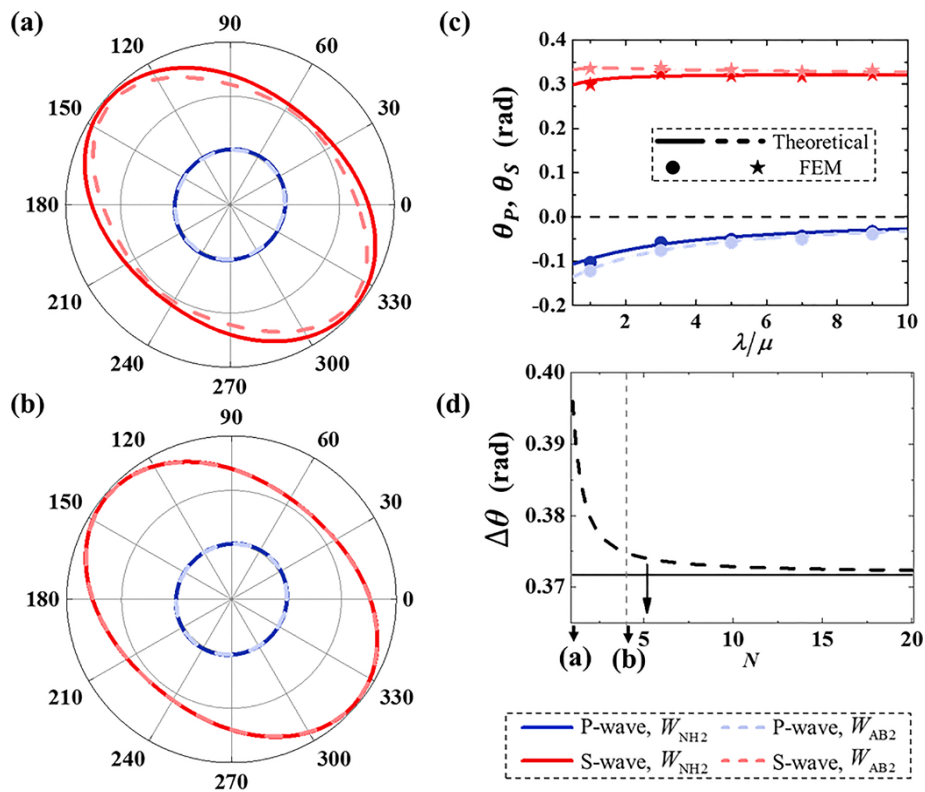

Fig.5

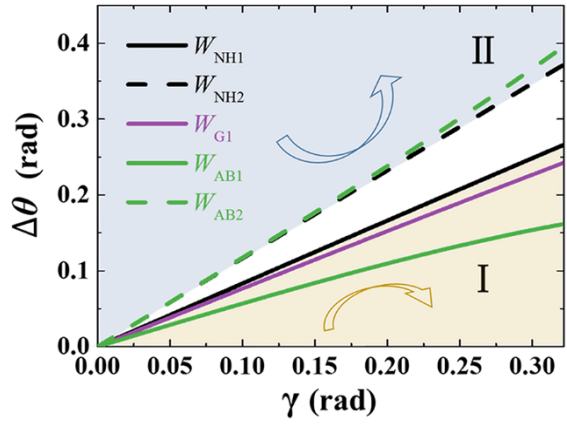

Fig.6

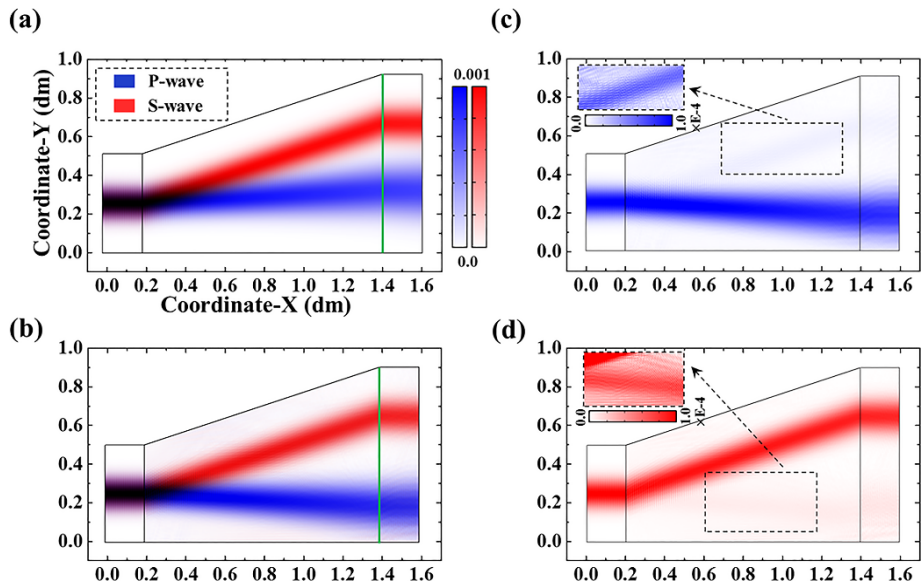

Fig.7

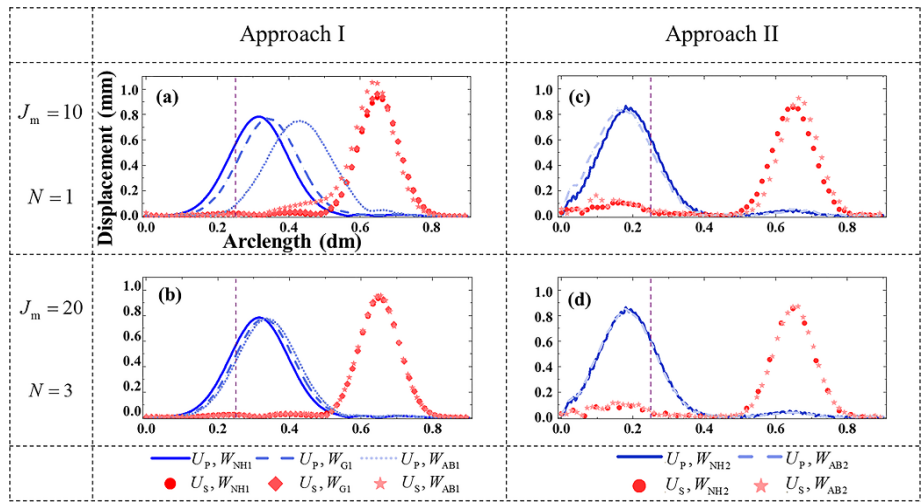